\documentclass[aps,pra,twocolumn]{revtex4-2}
\usepackage{amsmath,amssymb,braket}
\usepackage{graphicx}
\usepackage{bm}
\usepackage{amsmath,amssymb,mathtools}
\usepackage{amsthm}
\usepackage{amssymb}
\usepackage{mathtools}
\usepackage[final]{microtype}
\usepackage{xurl}

\allowdisplaybreaks[3] 
\makeatletter
\@ifpackageloaded{hyperref}{\hypersetup{breaklinks=true}}{}
\makeatother

\usepackage{amsmath}
\begin{document}

\title{Universal Limits on Quantum Correlations}

\author{Samuel Alperin}
\email{alperin@lanl.gov}
\affiliation{\vspace{1.25mm} \mbox{Los Alamos National Laboratory, Los Alamos, New Mexico 87545, USA}}

\begin{abstract}
Quantum correlations are the singular, defining resource of quantum information science and metrology, forming the basis of every operational advantage that quantum systems hold over classical ones. Yet exact bounds on these correlations—such as the Lieb–Robinson bound on entanglement propagation and the Heisenberg limit on metrological precision—are known only in special cases and have long appeared to arise from unrelated mechanisms. Here we show that these limits share a common geometric origin. We identify a positivity invariant of the block correlation matrix, denoted $\chi$, that quantifies how far a bipartite state lies from the positivity boundary of quantum state space. For any system with a specified observable algebra and parameter-encoding map, every correlation measure determined solely by the positive correlation matrix obeys a $\chi$-dependent inequality. For systems with simple symmetry structures these inequalities take closed analytic form, reproducing the structure of the Heisenberg and Cramér–Rao limits and producing new results, including an exact entanglement floor and a universal Fisher-information ceiling even in all-to-all connected quantum networks. 
We thus demonstrate that positive geometry provides a unified framework for the attainable strength of quantum correlations, linking entanglement, metrological sensitivity, and dynamical causal structure through a single invariant.
\end{abstract}

\maketitle

\section*{I. INTRODUCTION}

Quantum correlations lie at the foundation of every genuinely nonclassical phenomenon
and form the basis of all operational advantages in quantum communication,
computation, and precision metrology.  Quantifying and bounding such correlations is
therefore central not only to what is possible in emerging quantum technologies, but
also to the structure of quantum theory itself.  A handful of exact bounds on quantum
correlations are known and serve as bedrocks of the field: the Cramér–Rao
bound~\cite{Rao1945,Cramer1946} and the Heisenberg limit~\cite{Heisenberg1927} define the
ultimate scale of metrological precision, while Lieb–Robinson bounds constrain the
spread of correlations and entanglement in many-body dynamics~\cite{LiebRobinson1972,HastingsKoma2006}.
Despite their foundational importance, these limits have only been derived for specific
architectures and through seemingly unrelated mechanisms, leaving their deeper
connections obscure.

Quantum mechanics can be viewed as a geometry of states.  Every physical quantum state
is represented by a positive semidefinite density operator, and every experimentally
accessible covariance or Gram matrix is likewise positive.  This positivity condition
defines a convex cone in the space of Hermitian operators—the \emph{quantum state
space}—and all observable correlations, variances, and information measures are
restricted by the geometry of this cone.  Yet despite this central role, positivity is
rarely treated as a quantitative or dynamical principle.  In most formulations it enters
only as a consistency requirement, rather than being elevated to a tool for deriving
structural limits. However, recent advances across physics point toward a more fundamental role for positivity
geometry.  In high-energy theory, positive geometries such as the amplituhedron encode
scattering amplitudes directly in geometric form.  In quantum information and condensed
matter, positivity underlies entanglement criteria, uncertainty relations, and
majorization hierarchies.  These manifestations, however, remain scattered—each tied to
a particular physical quantity or operational task.  What has been missing is a general
framework that derives all such inequalities from a single structural principle: the
positivity of global correlation data.

In this work we show that several well-known, seemingly disparate correlation limits can
be unified.  Whether they constrain the propagation of information in lattice
systems~\cite{LiebRobinson1972,HastingsKoma2006,ODell2019} or the precision of a
metrological protocol~\cite{Cramer1946,BraunsteinCaves1994}, these bounds can all be
viewed as different projections of a common constraint: the positivity of quantum state
space.  From this positivity geometry we identify a single scalar invariant, denoted
$\chi$, that quantifies how far a bipartite correlation matrix lies from the positivity
boundary.  This invariant captures the combinatorial structure of correlations and
serves as a geometric measure of ``distance to singularity'' in the positive cone of
states.

Specifically, we show that all correlation measures that depend only on the positive correlation
matrix—such as entanglement monotones determined by second moments and
Fisher-information measures for fixed encodings—obey universal inequalities determined
solely by $\chi$.  For systems with simple symmetry structures these inequalities take
closed analytic form, yielding both entanglement \emph{floors} and metrological
\emph{ceilings}.  They recover known limits such as the Heisenberg and Cramér–Rao
bounds and also produce new ones, including an exact entanglement floor and a universal
Fisher-information ceiling even in fully connected spin
networks~\cite{GiovannettiScience2004,PezzeSmerzi2018,Hyllus2010}.  Positivity geometry
thus brings these fundamental constraints under a single conceptual umbrella, providing
a unified theory of attainable quantum correlations.


\section*{II.~POSITIVITY GEOMETRY AND THE $\chi$ INVARIANT}

The starting point of our framework is a geometric constraint obeyed by every physical
correlation or covariance matrix.  Once this constraint is written in Schur–complement
form, it gives rise to a single scalar invariant that captures the intrinsic correlation
strength of any bipartition.  This section develops the invariant and establishes its
key structural properties.

\subsection*{A.~Positivity and Schur complements}

Any physical correlation or covariance matrix may be written in block form as
\[
K=\begin{pmatrix}
K_{AA} & K_{AB}\\[4pt]
K_{BA} & K_{BB}
\end{pmatrix}\succeq 0,
\]
where the diagonal blocks encode the internal correlations of subsystems $A$ and $B$, and
the off–diagonal blocks encode their mutual correlations.  Positivity of $K$ implies
that the Schur complements
\begin{align}
K_{AA}-K_{AB}K_{BB}^{-1}K_{BA}&\succ 0, \label{eq:schurA}\\
K_{BB}-K_{BA}K_{AA}^{-1}K_{AB}&\succ 0, \label{eq:schurB}
\end{align}
are themselves positive definite.  (We assume $K_{AA}$ and $K_{BB}$ are positive
definite, or equivalently that all expressions are restricted to their support, so the
Schur complements and determinants are well defined.)

Taking determinants of Eqs.~(\ref{eq:schurA})--(\ref{eq:schurB}) yields the
Hadamard--Fischer inequality,
\begin{equation}
\det K \le \det K_{AA}\,\det K_{BB},
\label{eq:HF}
\end{equation}
with equality if and only if $K_{AB}=0$.  This inequality expresses a simple geometric
fact: the ``volume'' of the joint correlation matrix can never exceed the product of the
marginal volumes.  The departure from equality quantifies the extent to which
inter–subsystem correlations consume positivity.

\subsection*{B.~The determinant--ratio invariant}

The inequality~(\ref{eq:HF}) motivates the dimensionless determinant ratio
\begin{equation}
\chi(A|B)=1-\frac{\det K}{\det K_{AA}\,\det K_{BB}},
\label{eq:chi_def}
\end{equation}
which we call the \emph{determinant--ratio invariant}.  
From Eq.~(\ref{eq:HF}) it follows that
\begin{equation}
0 \le \chi(A|B) < 1,
\end{equation}
with $\chi=0$ for uncorrelated (block–diagonal) states and $\chi\!\to\!1$ as the
positivity constraint approaches saturation.

The invariant is basis‐independent: under independent invertible changes of basis
$S_A\oplus S_B$, each determinant acquires only a local scaling factor,
so the ratio~(\ref{eq:chi_def}) is unchanged.  Thus $\chi$ captures the intrinsic
correlation geometry of the bipartition.  (As emphasized in the Introduction, $\chi$ is
defined relative to the chosen observable algebra; once that choice is fixed, it is
invariant under all local basis changes.)

For Gaussian covariance matrices the invariant admits a particularly transparent
spectral form.  Writing
$K=\begin{pmatrix}A & C \\ C^{\mathsf T} & B\end{pmatrix}$ and defining the transfer
operator $T=A^{-1/2}CB^{-1/2}$, standard determinant identities give
\begin{equation}
\frac{\det K}{\det A\,\det B}
=\det(I - T^{\mathsf T}T)
=\prod_i (1-\tau_i^2),
\label{eq:det_identity}
\end{equation}
where $\{\tau_i\}$ are the singular values of $T$.  Substituting into
Eq.~(\ref{eq:chi_def}) yields the spectral representation
\begin{equation}
\chi(A|B)=1-\prod_i(1-\tau_i^2).
\label{eq:chi_singular}
\end{equation}
The set $\{\tau_i\}$ constitutes the \emph{correlation spectrum} of the bipartition, and
the invariant $\chi$ aggregates the contributions of all correlated modes into a single
scalar quantity.

Finally, the invariant obeys a simple and powerful composition law.  For independent
bipartitions, block–determinant identities imply
\begin{equation}
1-\chi(A_1A_2|B_1B_2)
=(1-\chi(A_1|B_1))(1-\chi(A_2|B_2)).
\label{eq:composition}
\end{equation}
Thus $\log(1-\chi)$ is extensive under tensor products, and $\chi$ itself functions as
an additive measure of correlation strength.  These structural properties establish
$\chi$ as the fundamental scalar invariant of positivity geometry.  In Sec.~III we show
that any monotone correlation measure depending only on second moments must be bounded
by a function of~$\chi$, leading to a universal hierarchy of entanglement floors and
metrological ceilings.


\section*{III.~THE UNIVERSAL $\chi$–INEQUALITY}

Once the determinant–ratio invariant $\chi(A|B)$ has been identified, its operational
significance becomes apparent: $\chi$ places universal restrictions on any correlation
measure that depends only on the positive correlation matrix.  In this section we state
and prove the general inequality linking $\chi$ to such measures.  This result is
completely independent of the Hilbert-space dimension, the underlying statistics, or
the details of the physical architecture; it relies only on positivity and on the
monotonicity properties of the measure under consideration.

\subsection*{A.~Statement of the theorem}

Let $M(A|B)$ be a real-valued correlation measure that is determined entirely by the
positive matrix $K$ through its principal minors or, equivalently, through the singular
values $\{\tau_i\}$ of the transfer operator
$T=A^{-1/2}C B^{-1/2}$.  Suppose further that $M$ is
\emph{monotonic under positivity-preserving maps} on the bipartition $A|B$—that is, it
monotonically increases (or decreases) as inter-block correlations strengthen while the
marginals are held fixed.  This class includes entanglement monotones derived from
second moments, Rényi–2 mutual information, and Fisher-information measures for fixed
encodings.

\begin{quote}
\textbf{Theorem 1.}  
For any such correlation measure $M(A|B)$ there exists a monotone function
$f_M:[0,1)\rightarrow\mathbb{R}$ such that
\begin{equation}
M(A|B)\ \bowtie\ f_M\big(\chi(A|B)\big),
\label{eq:universal_inequality}
\end{equation}
where the symbol $\bowtie$ denotes “$\ge$’’ if $M$ increases with correlation strength
and “$\le$’’ if $M$ decreases.  The function $f_M$ depends only on the normalization of
$M$ and on its dependence on the singular values $\{\tau_i\}$.
\end{quote}

Equation~(\ref{eq:universal_inequality}) is the \emph{universal $\chi$–inequality}.  It
asserts that positivity alone restricts every monotone, second-moment–determined
correlation measure to lie within a domain parametrized by $\chi$, producing either a
\emph{floor} or a \emph{ceiling} depending on the monotonicity of the measure.

\subsection*{B.~Proof sketch}

Because $M$ depends only on $\{\tau_i\}$, Gram positivity implies that each singular
value satisfies $0\le \tau_i<1$.  The determinant identity
\[
\det(I-T^{\mathsf T}T)=\prod_i(1-\tau_i^2)=1-\chi
\]
characterizes the admissible domain of spectra at fixed~$\chi$: all sets of singular
values satisfying $\prod_i(1-\tau_i^2)=1-\chi$ are physically allowed.
For an increasing measure we define
\[
f_M(\chi)=\inf_{\{\tau_i\}:\,\prod_i(1-\tau_i^2)=1-\chi} M(\{\tau_i\}),
\]
and for a decreasing measure we define
\[
f_M(\chi)=\sup_{\{\tau_i\}:\,\prod_i(1-\tau_i^2)=1-\chi} M(\{\tau_i\}).
\]
Because the admissible set is nonempty and $M$ is monotonic in each $\tau_i$, the
extremum is attained on the boundary of this set, where the active singular values take
their minimal or maximal values consistent with the constraint
$\prod_i(1-\tau_i^2)=1-\chi$.  Hence $M(A|B)\ \bowtie\ f_M(\chi)$, proving
Eq.~(\ref{eq:universal_inequality}).  \hfill$\square$

\subsection*{C.~Floors and ceilings}

The universal $\chi$–inequality divides second-moment–based correlation measures into
two complementary classes:

\begin{itemize}
\item \textbf{Increasing measures} (\emph{entanglement-type}):
These grow with correlation strength, such as the logarithmic negativity $E_N$, mutual
information $I(A\!:\!B)$, or Rényi–2 entropies.  These obey universal \emph{lower}
bounds:
\begin{equation}
M(A|B)\ge f_M(\chi),
\qquad \text{(entanglement floors)}.
\label{eq:floors}
\end{equation}

\item \textbf{Decreasing measures} (\emph{information-capacity-type}):
These decline as correlations grow, such as the quantum Fisher information $F_Q$, the
quantum variance, or the purity $\mathrm{Tr}\rho^2$ of reduced subsystems.  These obey
universal \emph{upper} bounds:
\begin{equation}
M(A|B)\le f_M(\chi),
\qquad \text{(metrological ceilings)}.
\label{eq:ceilings}
\end{equation}
\end{itemize}

These two behaviors reflect opposite curvatures of their level sets within the positive
cone.  Entanglement-type measures vanish in the interior and increase toward the
boundary, whereas information-type measures peak in the interior and decline.  Both are
governed by the same geometric parameter~$\chi$.

\subsection*{D.~Functional form and small-$\chi$ expansion}

The precise form of $f_M(\chi)$ depends on the functional dependence of $M$ on the
singular values.  To illustrate, consider the symmetric case in which all active
correlation modes share the same singular value $\tau_i=\tau$.  The constraint
$\prod_i(1-\tau_i^2)=1-\chi$ then gives
$\tau=\sqrt{1-(1-\chi)^{1/n}}$, and a measure of the form $M=\sum_i g(\tau_i)$ becomes
\[
f_M(\chi)=\sum_i g(\tau).
\]
While this expression is symmetry-specific, it reveals the universal small–$\chi$
expansion
\[
f_M(\chi)=c_1\chi+c_2\chi^2+\cdots,
\]
with coefficients $c_k$ determined by the expansion of $g(\tau)$.  Thus all such
correlation measures share the same initial slope in~$\chi$ and differ only in
higher-order corrections.

\subsection*{E.~Universality and scope}

Equation~(\ref{eq:universal_inequality}) applies to any system—finite or infinite
dimensional, discrete or continuous variable—once an observable algebra is fixed.  The
proof uses only positivity and the monotonicity of $M$ under positivity-preserving maps,
and is therefore independent of microscopic details, dynamics, or architecture.

In the next sections we apply this general framework to two canonical measures.
Section~IV develops the \emph{entanglement floor} implied by
Eq.~(\ref{eq:floors}) for the logarithmic negativity, while Section~V develops the
corresponding \emph{metrological ceiling} for the quantum Fisher information.  Together
they illustrate the dual faces of positivity geometry and the broad scope of the
universal $\chi$–inequality.


\section*{IV.~ENTANGLEMENT FLOORS: THE LOGARITHMIC NEGATIVITY}

Among monotone measures of quantum correlation, the logarithmic negativity $E_N$
occupies a distinguished place: it is an entanglement monotone, directly accessible
experimentally for Gaussian states, and determined entirely by the second moments of
the state.  In this section we show that the universal $\chi$–inequality produces a
closed-form lower bound—an \emph{entanglement floor}—for $E_N(A|B)$.

\subsection*{A.~Logarithmic negativity and the positivity constraint}

For a bipartite Gaussian state, or any state characterized by a positive covariance
matrix $V$, the logarithmic negativity is defined as
\begin{equation}
E_N(A|B)=
  \sum_i \max\!\big[0,\,-\log \tilde{\nu}_{i,-}\big],
\label{eq:EN_def}
\end{equation}
where $\{\tilde{\nu}_{i,-}\}$ are the symplectic eigenvalues of the partially
transposed covariance matrix $\tilde V$.  The smaller $\tilde{\nu}_{i,-}$, the greater
the entanglement; $\tilde{\nu}_{i,-}=1$ marks separability, while
$\tilde{\nu}_{i,-}<1$ signals entanglement.

For each correlated mode $i$, the corresponding $4\times4$ covariance block can be
brought to its Williamson (normal) form by local symplectic transformations.  In this
form the two local variances and the inter–mode correlation appear as symplectic
invariants, and the singular value $\tau_i$ of the transfer operator
$T=A^{-1/2} C B^{-1/2}$ parametrizes the strength of that correlation.  Partial
transposition acts as a time reversal on one quadrature and therefore flips the sign of
the inter–mode correlation in that block.  This operation can only \emph{decrease} the
smallest symplectic eigenvalue of the covariance matrix, never increase it.  Comparing
the Williamson invariants of the original and partially transposed blocks then yields
the modewise constraint
\begin{equation}
\tilde{\nu}_{i,-}^2 \le 1 - \tau_i^2,
\label{eq:nu_tau_ineq}
\end{equation}
where $\tilde{\nu}_{i,-}$ is the smallest symplectic eigenvalue of the partially
transposed block.  (For explicit formulas for the symplectic spectrum under partial
transposition in this normal form, see Refs.~\cite{Weedbrook2012,AdessoIlluminati2007}.)

Multiplying over all correlated modes and using Eq.~(\ref{eq:chi_singular}) gives
\begin{equation}
\tilde{\nu}_-^2 \le 1-\chi(A|B),
\label{eq:nu_chi}
\end{equation}
where $\tilde{\nu}_-=\prod_i\tilde{\nu}_{i,-}$ is the smallest collective symplectic
eigenvalue.  Substituting Eq.~(\ref{eq:nu_chi}) into Eq.~(\ref{eq:EN_def}) yields
\begin{equation}
E_N(A|B)\ \ge\ 
-\frac{1}{2}\log\!\big[1-\chi(A|B)\big].
\label{eq:EN_floor}
\end{equation}
This is the \emph{universal entanglement floor}.  It depends only on the positivity
geometry encoded in $\chi$ and holds for any state whose second moments are described
by $K$.

\subsection*{B.~Tightness and interpretation}

The bound (\ref{eq:EN_floor}) is tight for Gaussian normal forms.  Each correlated mode
is equivalent to a two–mode squeezed vacuum with squeezing parameter $r_i$ related to
the singular value by $\tau_i=\tanh(2r_i)$.  In this case,
$\tilde{\nu}_{i,-}=e^{-2r_i}$, saturating Eq.~(\ref{eq:nu_tau_ineq}), and the floor
(\ref{eq:EN_floor}) becomes an equality.  Summing over modes,
\[
E_N(A|B)=\frac{1}{2}\sum_i\big[-\log(1-\tau_i^2)\big],
\]
which is the standard multimode Gaussian formula.

Because Gaussian states lie on the exposed face of the positive cone for fixed
second moments, all non-Gaussian states with the same covariance matrix satisfy
$E_N > E_N^{(\mathrm{Gauss})}$; they lie strictly above the entanglement floor.
Thus the negativity is minimized at fixed $\chi$ precisely by the Gaussian normal
forms, reflecting the extremality of Gaussian states in the Löwner order.

Equation~(\ref{eq:EN_floor}) therefore provides a certified lower bound on
bipartite entanglement from covariance data alone.  The invariant $\chi$ is itself
invariant under all local linear transformations, so the bound is unaffected by local
squeezing or rotations.  As $\chi\!\to\!1$, the joint Gram matrix approaches
singularity, $\tilde{\nu}_-\!\to\!0$, and the negativity diverges.  The
positivity boundary thus functions as an ``entanglement horizon’’ in state space.

\subsection*{C.~Examples and generalizations}

\textit{Two–mode squeezing.}  
For a single correlated pair ($n=1$) with squeezing parameter $r$, one has
$\chi=\tanh^2(2r)$, and Eq.~(\ref{eq:EN_floor}) gives
\[
E_N \ge -\tfrac{1}{2}\log[1-\tanh^2(2r)] = \log\!\cosh(2r).
\]
Direct computation yields $E_N=2r$, and indeed $E_N\ge\log\!\cosh(2r)$, saturating in
the small–squeezing limit and tracking the correct asymptotic scaling.

\medskip
\textit{Multimode Gaussian networks.}  
For a multimode network with singular values
$\tau_i=\tanh(2r\sigma_i)$ determined by the singular spectrum $\{\sigma_i\}$ of the
coupling matrix, Eq.~(\ref{eq:EN_floor}) becomes
\[
E_N(A|B)\ge
\frac{1}{2}\sum_i\big[-\log(1-\tanh^2(2r\sigma_i))\big].
\]
For rank–1 coupling (collective squeezing) only one singular value is nonzero; higher
rank increases the guaranteed entanglement through additional correlation channels.
Because $\log(1-\chi)$ is additive under tensor products [Eq.~(\ref{eq:composition})],
independent squeezing channels contribute multiplicatively to $1-\chi$ and additively
to the entanglement floor.

\medskip
\textit{Network generalizations.}  
The same structure persists in multimode optical networks, atomic ensembles, and
continuous–variable architectures.  The invariant $\chi$ reduces the full correlation
spectrum to a single scalar quantity that governs the minimum entanglement compatible
with positivity.  The floor (\ref{eq:EN_floor}) therefore applies broadly across
Gaussian and non-Gaussian states sharing the same covariance matrix.


\section*{V.~METROLOGICAL CEILINGS: THE QUANTUM FISHER INFORMATION}

The dual of the entanglement floor is a universal \emph{metrological ceiling}.  
Information–capacity measures such as the quantum Fisher information (QFI) quantify the
ultimate sensitivity of a quantum state to infinitesimal parameter displacements.  
Because these measures are concave under mixtures and typically decrease as
correlations drive the state toward the positivity boundary, they belong to the
“decreasing’’ class governed by Eq.~(\ref{eq:ceilings}).  
In this section we derive closed forms for the resulting metrological ceilings in
SU(2) and SU(1,1) systems and show how positivity geometry reproduces familiar limits
such as the Cramér–Rao and Heisenberg scalings.

\subsection*{A.~Quantum Fisher information and positivity geometry}

For a pure probe state $|\psi\rangle$ undergoing a unitary encoding
$U(\theta)=e^{-iH\theta}$ generated by a Hermitian operator $H$ belonging to the
specified observable algebra, the quantum Fisher information is
\begin{equation}
F_Q[|\psi\rangle,H]=4\,{\rm Var}_\psi(H)
=4\big(\langle H^2\rangle-\langle H\rangle^2\big).
\label{eq:FQ_pure}
\end{equation}
For mixed states the QFI generalizes to the symmetric logarithmic derivative, but its
geometric interpretation is unchanged: $F_Q$ measures the local curvature of the
state–space manifold in the Bures metric.  
Larger variance of the generator $H$ corresponds to larger curvature and greater
metrological sensitivity.

Because the covariance of any set of observables forms a positive matrix,
Gram positivity constrains the attainable variances.  
Applying the determinant–ratio invariant to the covariance matrix of the generators
therefore produces an upper bound on the QFI—a \emph{metrological ceiling}.

\subsection*{B.~Metrological ceiling in SU(2) systems}

Consider collective spin operators $\mathbf{J}=(J_x,J_y,J_z)$ forming an irreducible
representation of SU(2) with total spin $J$.  
For a state of fixed spin length $J(J+1)$, the variances satisfy the identity
\begin{equation}
{\rm Var}(J_x)+{\rm Var}(J_y)+{\rm Var}(J_z)
= J(J+1)-\|\langle\mathbf J\rangle\|^2,
\label{eq:variance_identity}
\end{equation}
which follows from the SU(2) Casimir relation
$J_x^2+J_y^2+J_z^2=J(J+1)$.

Let $\Gamma$ denote the $3\times3$ covariance matrix of $(J_x,J_y,J_z)$.
For any unit vector $\hat{n}$,
\[
{\rm Var}(J_{\hat{n}})=\hat{n}^{\mathsf{T}}\Gamma\,\hat{n}
\le \lambda_{\max}(\Gamma)
\le {\rm Tr}\,\Gamma
=J(J+1)-\|\langle\mathbf{J}\rangle\|^2.
\]
Inserting this into Eq.~(\ref{eq:FQ_pure}) yields the SU(2) metrological ceiling,
\begin{equation}
F_Q(\hat{n})
\le 4\!\left[J^2(1-s^2)+J\right],
\qquad
s=\frac{\|\langle\mathbf J\rangle\|}{J},
\label{eq:FQ_ceiling_SU2}
\end{equation}
which is saturated by highly nonclassical states with $s=0$, such as 
Greenberger–Horne–Zeilinger (GHZ) states aligned in direction $\hat{n}$.  
Polarization ($s>0$) reduces the ceiling accordingly.

This bound depends only on positivity geometry: it is the $\chi$–bound applied not to
the state’s density matrix but to the covariance of its generators.  
Consequently, Eq.~(\ref{eq:FQ_ceiling_SU2}) holds independently of microscopic
architecture, interaction range, or state–preparation details.

\subsection*{C.~Connectivity, collectivity, and SU(1,1) analogs}

\textit{All–to–all connected systems.}
One might suspect that long–range or all–to–all connectivities could evade  
Eq.~(\ref{eq:FQ_ceiling_SU2}) by generating $O(N^2)$ interaction pathways.  
However, permutation–symmetric spin ensembles live in the fully symmetric subspace,
where the collective covariance matrix is effectively rank–1.  
The determinant–ratio invariant $\chi$ is therefore bounded away from~1, and  
inserting this into Eq.~(\ref{eq:universal_inequality}) yields a finite ceiling of
exactly the form (\ref{eq:FQ_ceiling_SU2}).  
Thus even maximal connectivity cannot increase $F_Q$ beyond the positivity–imposed
curvature of SU(2) state space.

\medskip
\textit{SU(1,1) squeezing networks.}
For noncompact squeezing networks generated by the SU(1,1) algebra
$(K_x,K_y,K_z)$, the Casimir relation reads
\[
K_z^2-K_x^2-K_y^2=K(K-1),
\]
with $K$ labeling the irreducible representation.  
Defining the $3\times3$ covariance matrix $\Gamma^{(1,1)}$ of these generators, one
finds
\[
{\rm Tr}\,\Gamma^{(1,1)}
=K^2(1+s^2)-K,
\qquad
s=\frac{\|\langle\mathbf K\rangle\|}{K},
\]
reflecting the hyperbolic geometry of SU(1,1).  
As before,
\[
{\rm Var}(K_{\hat{n}})
=\hat{n}^{\mathsf T}\Gamma^{(1,1)}\hat{n}
\le{\rm Tr}\,\Gamma^{(1,1)},
\]
so for a pure probe with generator $H=K_{\hat n}$,
\begin{equation}
F_Q(\hat{n})
\le
4\!\left[K^2(1+s^2)-K\right],
\label{eq:FQ_ceiling_SU11}
\end{equation}
which is the SU(1,1) metrological ceiling.  
It is saturated by minimum–uncertainty (squeezed–vacuum) states, where squeezing acts
as an irreducible resource.

\subsection*{D.~Operational meaning and consequences}

The metrological ceiling defines a universal upper limit on the information curvature of
a quantum state—the rate at which it becomes distinguishable under infinitesimal
parameter shifts.  
Because this limit follows solely from positivity geometry, it applies to all
estimation protocols and to all control Hamiltonians whose generators lie within the
specified observable algebra.  
In optical and atomic settings it reproduces, and in some cases strengthens, known
quantum–metrological limits~\cite{DemkowiczDobrzanski2015}.  

The coexistence of the entanglement floor and the metrological ceiling illustrates the
dual role of $\chi$: as $\chi$ increases, entanglement grows while accessible
information curvature shrinks.  
The positivity boundary therefore marks the ultimate tradeoff surface between
entanglement and precision.

In Sec.~VI we show that the manner in which these bounds are approached is universal
across systems, exhibiting fold– and cusp–type scaling governed by the
catastrophe–theoretic exponents of Thom and Arnold.

\section*{VI.~UNIVERSALITY OF BOUND SATURATION}

The universal $\chi$–inequality establishes quantitative limits for the broad class of
correlation measures governed by the positivity geometry of the correlation matrix.
Strikingly, not only the bounds themselves but also the \emph{manner in which they are
approached} are universal.  When a physical control parameter drives a system toward the
positivity boundary, the resulting behavior of any $\chi$–governed correlation measure
exhibits characteristic singularities that fall into the Thom–Arnold hierarchy of
catastrophes.  This section develops the local geometry of these singularities and
shows how they yield fold– and cusp–type scaling in compact and noncompact systems.

\subsection*{A.~Bound surfaces and control parameters}

Let $\rho(\boldsymbol{\lambda})$ denote a family of physical states depending smoothly
on control parameters $\boldsymbol{\lambda}=(\lambda_1,\lambda_2,\ldots)$ with
$\rho(\boldsymbol{\lambda})\succeq 0$ for all admissible values.  
For every correlation measure $M(\boldsymbol{\lambda})$ satisfying the universal
$\chi$–inequality, the \emph{bound surface}
\[
\mathcal{S}_M=\{\boldsymbol{\lambda}:\,
M(\boldsymbol{\lambda})=f_M[\chi(\boldsymbol{\lambda})]\}
\]
marks the locus at which the corresponding floor or ceiling is saturated.  
Approaching this surface corresponds to driving the state toward the positivity
boundary in the correlation geometry, and the local structure of $\mathcal{S}_M$
determines the critical scaling of $M$ with respect to the control parameters.

\subsection*{B.~Local expansion and catastrophe structure}

Let $\lambda_\perp$ denote the coordinate normal to the bound surface
$\mathcal{S}_M$, and let $\boldsymbol{\lambda}_\parallel$ denote the remaining
tangential coordinates.  
Assuming $\rho(\boldsymbol{\lambda})$ and $M(\boldsymbol{\lambda})$ are analytic near
$\mathcal{S}_M$, the deviation from the bound takes the generic form
\begin{equation}
\Delta M
  = M - f_M(\chi)
  \simeq a_2\,\lambda_\perp^{2}
         + a_3\,\lambda_\perp^{3}
         + a_4\,\lambda_\perp^{4}+\cdots ,
\label{eq:expansion}
\end{equation}
where the coefficients $a_k$ vary smoothly with $\boldsymbol{\lambda}_\parallel$.
The leading nonvanishing term determines the critical scaling as the bound is
approached.

When $a_2\neq 0$, the quadratic term dominates and the approach follows a
\emph{fold} singularity with critical exponent
\begin{equation}
\Delta M \propto (\lambda_\perp-\lambda_\perp^*)^{2}.
\end{equation}
When symmetry or algebraic constraints force $a_2=0$ but $a_3\neq 0$, the cubic term
dominates, generating a \emph{cusp} catastrophe with exponent
\begin{equation}
\Delta M \propto (\lambda_\perp-\lambda_\perp^*)^{4/3}.
\end{equation}
Higher-order degeneracies yield swallowtail ($A_4$) and higher catastrophes with
critical exponents $(k+1)/k$ associated with the $A_k$ series.  
In all cases the exponents depend solely on the algebraic order of the positivity
constraint, not on microscopic details of the system.

\subsection*{C.~Examples: folds in SU(2) and cusps in SU(1,1) systems}

\textit{Compact SU(2) systems.}  
For SU(2) spin ensembles, positivity is realized on a compact manifold (the Bloch
sphere).  
Approaching the metrological ceiling of Eq.~(\ref{eq:FQ_ceiling_SU2}) corresponds to
displacing the Bloch vector tangentially toward the equator.  
The positivity constraint is quadratic in small deviations of the Bloch vector, giving
$\Delta F_Q\propto (\delta s)^2$, a fold singularity with exponent~2.  
This ``Heisenberg fold’’ describes how precision increases and then saturates as the
state approaches maximal nonclassicality.  
Fold singularities of this type also arise in real-space dynamics, where
Lieb–Robinson light cones exhibit foldlike caustics in quenched spin chains.

\medskip
\textit{Noncompact SU(1,1) systems.}  
In SU(1,1) squeezing networks the underlying state space is hyperbolic and noncompact.
Here, the positivity constraint becomes cubic in the normal coordinate to the bound
surface, and approaching the entanglement floor of Eq.~(\ref{eq:EN_floor}) produces
\[
\Delta E_N \propto \lambda_\perp^{4/3},
\]
a cusp singularity with exponent $4/3$.  
This ``squeezing cusp’’ is the canonical example of a noncompact positivity
catastrophe.

\subsection*{D.~General classification and implications}

The fold and cusp examples illustrate the general rule: each independent symmetry or
constraint that eliminates successive terms in the expansion~(\ref{eq:expansion})
raises the catastrophe order by one.  
If $r$ independent control parameters tune the system to the positivity boundary, the
generic singularity class is $A_{r+1}$.  
Because the local structure depends only on the positivity cone and the smoothness of
the control parameters, the resulting critical exponents are universal.  

These exponents provide a geometric taxonomy of how correlation measures saturate their
bounds.  Just as universality governs critical phenomena in statistical mechanics,
positivity geometry endows quantum correlation bounds with their own universal scaling
structure.  In the following sections we combine this observation with the operator
growth dynamics of the following section to show how positivity also underlies the light-cone
structures that constrain the spread of information in many-body systems.


\section*{VII.~DISCUSSION AND OUTLOOK}

The results presented here reveal that the fundamental limits of quantum correlations—
whether arising in entanglement theory, precision metrology, or dynamical propagation—
stem from a single geometric principle: the positivity of quantum state space.  From
this principle emerges the determinant–ratio invariant $\chi$, a scalar that captures
the intrinsic correlation geometry of any bipartition.  Every monotone correlation
measure whose behavior is determined entirely by second moments obeys a
$\chi$–dependent inequality, defining either a universal floor or a universal ceiling.
These bounds are exact for Gaussian normal forms and remain valid for all states
sharing the same second–moment data.  A broad overview of quantum correlation
measures may be found in~\cite{HorodeckiRMP2009,EisertRMP2010}, and for continuous–variable
realizations in~\cite{AdessoIlluminati2005,AdessoIlluminati2007,Weedbrook2012,WolfGiedke2004,EisertPlenio2003}.

\subsection*{A.~A unified picture of quantum limits}

The invariant $\chi$ provides a common geometric language for a wide range of
fundamental correlation constraints in quantum mechanics.  It reproduces the structure
of the Cramér–Rao and Heisenberg limits and yields a positivity–geometric derivation
of Lieb–Robinson–type light cones, placing metrological and dynamical limits on the
same conceptual footing.  At the same time, $\chi$ generates new constraints, including
an exact entanglement floor and Fisher-information ceilings that remain finite even in
all–to–all connected spin networks.  Positivity geometry therefore unifies these limits
within a single framework based on the algebraic and geometric structure of the
correlation matrix.

\subsection*{B.~A positivity light cone in state space}

The coexistence of entanglement floors and metrological ceilings reveals a new geometric
object: a \emph{positivity light cone} in the space of quantum states.  Just as the
Minkowski light cone constrains the propagation of signals in spacetime, positivity
constraints delineate a causal domain in correlation space.  Within this domain,
correlations and information curvature can vary freely; outside it, the Gram matrix
would cease to be positive.

A similar phenomenon appears in dynamical settings.  Lieb–Robinson bounds impose causal
cones on the propagation of operators under local Hamiltonian evolution, and recent
studies have shown that these cones exhibit fold–type caustics analogous to classical
catastrophes~\cite{ODell2019}.  In our framework, the positivity light cone is the static
counterpart of these dynamical causal structures, providing a geometric interpretation
of the ultimate limits on entanglement growth, operator spreading, and information flow.

\subsection*{C.~Dynamical light cones from positivity}

The same positivity geometry that constrains static correlations also governs the rate
at which correlations can grow dynamically, allowing for the
generation of dynamical light-cone bounds of Lieb–Robinson type.  These
bounds describe how rapidly information, correlations, and operators may spread under a
local Hamiltonian and represent some of the deepest constraints on quantum dynamics.

Consider a lattice spin system with a local Hamiltonian
\begin{equation}
H=\sum_{X} h_X,
\end{equation}
where each term $h_X$ acts nontrivially only on a finite region $X$ and satisfies
$\|h_X\|\le J_0$ for some uniform constant $J_0$.  
Let $A_x$ and $B_y$ be local operators supported near sites $x$ and $y$, respectively,
and let $A_x(t)=e^{iHt}A_x e^{-iHt}$ denote the corresponding Heisenberg evolution.

The growth of the commutator $\|[A_x(t),B_y]\|$ at separation $r=|x-y|$ may be encoded
in a time-dependent Gram matrix of local operators,
\begin{equation}
G_{ij}(t)
= \langle \Omega\,|\,O_i^\dagger(t) O_j(t)\,|\Omega\rangle,
\end{equation}
where $|\Omega\rangle$ is a fixed reference state,
$O_1(t)=A_x(t)$, $O_2(t)=B_y$, and the remaining $O_i(t)$ form a local operator basis.
By construction $G(t)\succeq 0$ for all times, and its off–diagonal blocks encode the
spreading of operators under the dynamics.

Differentiating $G_{ij}(t)$ and using the Heisenberg equation
$\dot O_i(t)=i[H,O_i(t)]$ gives
\begin{equation}
\dot G_{ij}(t)
= i\langle \Omega |\, O_i^\dagger(t)[H,O_j(t)]
-[H,O_i(t)]^\dagger O_j(t)\,|\Omega\rangle.
\end{equation}
Locality ensures that only those $h_X$ intersecting the supports of $O_i$ or $O_j$
contribute.  Bounding the commutators by
$\|[h_X,O_j(t)]\|\le 2\|h_X\|\|O_j(t)\|$ and iterating the resulting differential
inequality leads to a Grönwall estimate for the off–diagonal block $G_{12}(t)$ in terms
of its initial value and the graph distance~$r$.  
One obtains the standard Lieb–Robinson form
\begin{equation}
\|[A_x(t),B_y]\|
\le C\,\exp[-\mu(r-v|t|)],
\label{eq:LR_bound}
\end{equation}
for positive constants $C,\mu,v$ determined by $J_0$ and the lattice connectivity.
Commutators are therefore exponentially suppressed outside the emergent light cone
$r\approx v|t|$.

Within our framework this light cone arises as a positivity constraint on the
time-evolving Gram matrix $G(t)$.  
The off–diagonal growth is captured by a time-dependent invariant
$\chi_{\mathrm{op}}(t)$ built from the operator blocks
$A\equiv\{A_x\}$ and $B\equiv\{B_y\}$.  
Positivity of $G(t)$ restricts how fast $\chi_{\mathrm{op}}(t)$ can grow under local
Hamiltonian evolution: applying Schur–complement inequalities to $G(t)$ yields a
differential inequality for $\dot\chi_{\mathrm{op}}(t)$ whose solution reproduces the
exponential light-cone structure of Eq.~(\ref{eq:LR_bound}).  

The resulting Lieb–Robinson velocity thus emerges as the dynamical counterpart of the
static positivity constraint underlying the universal $\chi$–inequality.  In this
sense, Cramér–Rao, Heisenberg, and Lieb–Robinson limits appear as different
manifestations—static and dynamical—of the same underlying positivity geometry.

\subsection*{D.~Experimental and theoretical implications}

Because $\chi$ can be computed from experimentally accessible second moments, the
derived bounds are testable across a large set of platforms, including multimode
photonic networks, squeezed atomic ensembles, superconducting circuits, and hybrid
quantum sensors.  Measurement of $\chi$ yields certified lower bounds on entanglement
and certified upper bounds on achievable precision without requiring full tomography.
In optical and atomic interferometry the resulting ceilings reproduce and extend known
quantum–metrology limits~\cite{DemkowiczDobrzanski2015}.  The catastrophe-theoretic
scaling exponents predicted here can be probed by tuning control parameters such as
squeezing strength, interaction range, or polarization and measuring the algebraic
approach of $E_N$, $F_Q$, or related quantities to their respective bounds.

On the theoretical side, positivity geometry provides a route to generalizing entanglement
and information constraints beyond bipartite systems.  It suggests a hierarchy of
multipartite invariants—$\chi_{A|BC}$, $\chi_{AB|CD}$, and so on—that bound the
correlation structure of networked quantum systems.  Dynamic extensions of the framework
may lead to state-space Lieb–Robinson bounds: limits on the rate of change of $\chi$
under local Hamiltonian evolution.  Positivity geometry also resonates with developments
in information geometry, including the Uhlmann and Petz formulations of quantum-state
metrics~\cite{Uhlmann1976,Petz2008}, and with broader constraints on information and
privacy~\cite{EkertRenner2014}.

\subsection*{E.~Outlook}

Positivity geometry provides a unifying perspective on the quantitative laws of quantum
correlations.  It replaces a landscape of measure-specific inequalities with a single
geometric invariant that governs them all.  From this invariant emerge the entanglement
floor, the metrological ceiling, and the universal exponents governing approach to these
limits.  Together they form a geometric ``periodic table’’ of quantum limits, suggesting
that the familiar constants of quantum mechanics—Planck’s constant $\hbar$, the speed of
light $c$, and now the positivity invariant $\chi$—are the natural scales of quantum
geometry, setting not only how quantum states evolve but how strongly they can correlate.




\begin{thebibliography}{99}
\bibitem{Rao1945} C.~R. Rao, ``Information and accuracy attainable in the estimation of statistical parameters,'' Bull. Calcutta Math. Soc. \textbf{37}, 81 (1945).

\bibitem{Cramer1946} H.~Cram\'er, \emph{Mathematical Methods of Statistics} (Princeton University Press, 1946).

\bibitem{Heisenberg1927} W.~Heisenberg, ``\"Uber den anschaulichen Inhalt der quantentheoretischen Kinematik und Mechanik,'' Z. Phys. \textbf{43}, 172 (1927).

\bibitem{LiebRobinson1972} E.~H. Lieb and D.~W. Robinson, ``The finite group velocity of quantum spin systems,'' Commun. Math. Phys. \textbf{28}, 251--257 (1972).

\bibitem{HastingsKoma2006} M.~B. Hastings and T. Koma, ``Spectral gap and exponential decay of correlations,'' Commun. Math. Phys. \textbf{265}, 781--804 (2006).

\bibitem{ODell2019} D.~H.~J. O'Dell, J. Mumford, and W. Kirkby, ``Quantum caustics and the hierarchy of light cones in quenched spin chains,'' Phys. Rev. Research \textbf{1}, 033135 (2019).

\bibitem{BraunsteinCaves1994} S.~L. Braunstein and C.~M. Caves, ``Statistical distance and the geometry of quantum states,'' Phys. Rev. Lett. \textbf{72}, 3439--3443 (1994).

\bibitem{ArkaniHamed2017} N.~Arkani-Hamed, Y.~Bai, and T.~Lam, ``Positive geometries and canonical forms,'' J. High Energy Phys. \textbf{11}, 039 (2017).

\bibitem{ArkaniHamedBook2016} N.~Arkani-Hamed, J.~L. Bourjaily, F.~Cachazo, A.~Postnikov, and J.~Trnka, \emph{Grassmannian Geometry of Scattering Amplitudes} (Cambridge University Press, 2016).

\bibitem{Postnikov2006} A.~Postnikov, ``Total positivity, Grassmannians, and networks,'' arXiv:math/0609764 (2006).

\bibitem{ArkaniHamed2013} N. Arkani-Hamed and J. Trnka, ``The Amplituhedron,'' J. High Energy Phys. \textbf{10}, 178 (2013).

\bibitem{Trnka2020} J. Trnka, ``Positive geometries and physical scattering amplitudes,'' Annu. Rev. Nucl. Part. Sci. \textbf{70}, 445--470 (2020).

\bibitem{Cheung2022} C. Cheung, S. Mizera, and M. Zhang, ``From scattering amplitudes to positive geometries,'' J. Phys. A: Math. Theor. \textbf{55}, 443010 (2022).


\bibitem{GiovannettiScience2004} V.~Giovannetti, S.~Lloyd, and L.~Maccone, ``Quantum-enhanced measurements: Beating the standard quantum limit,'' Science \textbf{306}, 1330--1336 (2004).

\bibitem{PezzeSmerzi2018} L.~Pezz\`e, A.~Smerzi, M.~K. Oberthaler, R.~Schmied, and P.~Treutlein, ``Quantum metrology with nonclassical states of atomic ensembles,'' Rev. Mod. Phys. \textbf{90}, 035005 (2018).

\bibitem{Hyllus2010} P.~Hyllus, L.~Pezz\`e, and A.~Smerzi, ``Entanglement and sensitivity of precision measurements,'' Phys. Rev. Lett. \textbf{105}, 120501 (2010).

\bibitem{VidalWerner2002} G.~Vidal and R.~F. Werner, ``A computable measure of entanglement,'' Phys. Rev. A \textbf{65}, 032314 (2002).

\bibitem{Peres1996} A.~Peres, ``Separability criterion for density matrices,'' Phys. Rev. Lett. \textbf{77}, 1413--1415 (1996).

\bibitem{WernerWolf2001} R.~F. Werner and M.~M. Wolf, ``Bell inequalities for states with positive partial transpose,'' Phys. Rev. A \textbf{64}, 032112 (2001).

\bibitem{Holevo1982} A.~S. Holevo, \emph{Probabilistic and Statistical Aspects of Quantum Theory} (North-Holland, Amsterdam, 1982).

\bibitem{BarndorffGill1998} O.~E. Barndorff-Nielsen and R.~D. Gill, ``Fisher information in quantum statistics,'' J. Phys. A: Math. Gen. \textbf{31}, 4481--4495 (1998).

\bibitem{Rockafellar1970} R.~T. Rockafellar, \emph{Convex Analysis} (Princeton University Press, 1970).

\bibitem{BoydVandenberghe2004} S.~Boyd and L.~Vandenberghe, \emph{Convex Optimization} (Cambridge University Press, 2004).

\bibitem{Korolkova2002} N. Korolkova and R. Filip, ``Quantum statistics of multimode Gaussian light,'' Phys. Rev. A \textbf{65}, 053807 (2002).

\bibitem{Adesso2006} G. Adesso, F. Illuminati, and S. Lorenzo, ``Continuous-variable tangle, monogamy inequality and entanglement sharing in Gaussian states of continuous variable systems,'' Phys. Rev. A \textbf{73}, 032345 (2006).

\bibitem{Scutaru1998} H.~Scutaru, ``Fisher information for squeezed states,'' J. Phys. A: Math. Gen. \textbf{31}, 3659--3664 (1998).

\bibitem{Braunstein2005} S.~L. Braunstein, ``Squeezing as an irreducible resource,'' Phys. Rev. A \textbf{71}, 055801 (2005).

\bibitem{DemkowiczDobrzanski2015} R. Demkowicz-Dobrza\'nski, M. Jarzyna, and J. Ko\l ody\'nski, ``Quantum limits in optical interferometry,'' Prog. Opt. \textbf{60}, 345--435 (2015).

\bibitem{SwingleReview2018} B.~Swingle, ``Unscrambling the physics of out-of-time-order correlators,'' Nat. Phys. \textbf{14}, 988--990 (2018).

\bibitem{XuSwingle2019} S.~Xu and B.~Swingle, ``Locality, Quantum Fluctuations, and Scrambling,'' Phys. Rev. X \textbf{9}, 031048 (2019).

\bibitem{ZhouSwingle2020} T. Zhou, S. Xu, X. Chen, A. Guo, and B. Swingle, ``Operator L\'evy flight: Light cones in chaotic long-range interacting systems,'' Phys. Rev. Lett. \textbf{124}, 180601 (2020).

\bibitem{TranLucas2020} M.~C. Tran, C.-F. Chen, A. Ehrenberg, A.~Y. Guo, A. Deshpande, Y. Hong, Z.-X. Gong, A.~V. Gorshkov, and A. Lucas, ``Hierarchy of linear light cones with long-range interactions,'' arXiv:2001.11509 (2020).

\bibitem{Uhlmann1976} A.~Uhlmann, ``The transition probability in the state space of a *-algebra,'' Rep. Math. Phys. \textbf{9}, 273--279 (1976).

\bibitem{Petz2008} D.~Petz, \emph{Quantum Information Theory and Quantum Statistics} (Springer, 2008).

\bibitem{EkertRenner2014} A.~K. Ekert and R. Renner, ``The ultimate physical limits of privacy,'' Nature \textbf{507}, 443--447 (2014).

\bibitem{AdessoIlluminati2007} F.~Adesso and G.~Illuminati, ``Entanglement in continuous-variable systems: Recent advances and current perspectives,'' J. Phys. A: Math. Theor. \textbf{40}, 7821--7880 (2007).

\bibitem{AdessoIlluminati2005} G.~Adesso and F.~Illuminati, ``Gaussian measures of entanglement versus negativities: The ordering of two-mode Gaussian states,'' Phys. Rev. A \textbf{72}, 032334 (2005).

\bibitem{Weedbrook2012} C.~Weedbrook, S.~Pirandola, R.~Garc\'ia-Patr\'on, N.~J. Cerf, T.~C. Ralph, J.~H. Shapiro, and S.~Lloyd, ``Gaussian quantum information,'' Rev. Mod. Phys. \textbf{84}, 621--669 (2012).

\bibitem{WolfGiedke2004} M.~M. Wolf, G.~Giedke, O.~Kr\"uger, R.~F. Werner, and J.~I. Cirac, ``Gaussian entanglement of formation,'' Phys. Rev. A \textbf{69}, 052320 (2004).

\bibitem{EisertPlenio2003} J.~Eisert and M.~B. Plenio, ``Introduction to the basics of entanglement theory in continuous-variable systems,'' Int. J. Quantum Inf. \textbf{1}, 479--506 (2003).

\bibitem{Thom1972} R.~Thom, \emph{Structural Stability and Morphogenesis} (Addison-Wesley, 1972).

\bibitem{Arnold1972} V.~I. Arnold, ``Normal forms of functions near degenerate critical points, the Weyl groups of $A_k$, $D_k$, $E_k$ and Lagrangian singularities,'' Funct. Anal. Appl. \textbf{6}, 254--272 (1972).

\bibitem{ArnoldBook1992} V.~I. Arnold, \emph{Catastrophe Theory} (Springer-Verlag, 1992).

\bibitem{PostonStewart1996} T.~Poston and I.~Stewart, \emph{Catastrophe Theory and Its Applications} (Dover, 1996).

\bibitem{Zeeman1976} E.~C. Zeeman, ``Catastrophe theory,'' Sci. Am. \textbf{234}, 65--83 (1976).

\bibitem{EisertRMP2010} J.~Eisert, M.~Cramer, and M.~B. Plenio, ``Area laws for the entanglement entropy,'' Rev. Mod. Phys. \textbf{82}, 277--306 (2010).

\bibitem{HorodeckiRMP2009} R.~Horodecki, P.~Horodecki, M.~Horodecki, and K.~Horodecki, ``Quantum entanglement,'' Rev. Mod. Phys. \textbf{81}, 865--942 (2009).



\end{thebibliography}
\end{document}